\documentclass[prl,aps,twocolumn,nofootinbib,10pt, showpacs]{revtex4-1}
\usepackage{amssymb}
\usepackage{amsmath}
\usepackage{appendix}
\usepackage{times}
\usepackage{graphicx}
\newcommand{\ket}[1]{|#1\rangle}
\newcommand{\bra}[1]{\langle #1|}
\usepackage{color}

\begin{document}

\title {Quantum spin Hall effect and topological phase transitions in honeycomb antiferromagnets}

\author{C. Niu}
\email{c.niu@fz-juelich.de}
\author{J.-P. Hanke}
\author{P. M. Buhl}
\author{G. Bihlmayer}
\author{D. Wortmann} 
\author{S. Bl\"{u}gel}
\author{Y. Mokrousov}
\affiliation
{Peter Gr\"{u}nberg Institut and Institute for Advanced Simulation, Forschungszentrum J\"{u}lich and JARA, 52425 J\"{u}lich, Germany
}	

\begin{abstract}
While the quantum spin Hall (QSH) effect and antiferromagnetic order constitute two of the most promising phenomena for embedding basic spintronic concepts into future technologies, almost all of the QSH insulators known to date are non-magnetic. Here, based on tight-binding arguments and first-principles theory, we predict two-dimensional antiferromagnets with honeycomb lattice structure to exhibit the QSH effect due to the combined symmetry of time reversal and spatial inversion. We identify functionalized Sn films as experimentally feasible examples which reveal  large band gaps rendering these systems ideal for energy efficient spintronics applications. Remarkably, we discover that tensile strain can tune the magnetic order in these materials, accompanied by a topological phase transition from the QSH to the quantum anomalous Hall phase. 
\end{abstract}

\maketitle
\date{\today}

During the past decade, research on topological insulators (TIs) became a rapidly-growing and attractive field, owing to their intriguing prospects, e.g., for dissipationless spintronics and quantum information~\cite{Hasan,Qi1}. Initially proposed in two dimensions, one hallmark of TIs is the quantum spin Hall (QSH) effect, which manifests in the quantization of the spin Hall conductivity and the emergence of helical edge states in the bulk band gap~\cite{Kane,Bernevig}. While these conducting edge states are protected by time-reversal symmetry, the coexistence of the TI phase with seemingly incompatible magnetic perturbations provides an enriched platform for exploring diverse topological phases in magnets. One example is the quantum anomalous Hall (QAH) effect~\cite{Haldane} where one pair of conducting edge states is suppressed as a consequence of the induced ferromagnetic (FM) order~\cite{Qi}. Only recently, this topologically non-trivial phase was observed experimentally in a magnetically doped TI~\cite{chang}, and several two-dimensional QAH insulators based on TIs are suggested theoretically~\cite{liu,yu,Qiao,zhanghb,wang,hzhang,wu,niubi}. 
Antiferromagnetism, on the other hand, provides a distinct flavor of magnetic order, which is perceived to hold bright promises for the efficient generation, detection, and transmission of spin currents, thereby launching off the field of antiferromagnetic (AFM) spintronics~\cite{Jungwirth}. 
Notably, recent works on non-trivial insulators~\cite{Mong,CFang,PZhou,JZhou}, topological semimetals~\cite{Tang}, and superconductors~\cite{ZWang} pioneered the QSH effect associated with antiferromagnetic order on a square lattice.

Understanding and utilizing the correlation between magnetism and topology is a key to further advances in topological antiferromagnetic  spintronics~\cite{Hasan,Qi1,Jungwirth}. This particularly 
concerns the design and exploitation of magnetic phase transitions for the purpose of shaping global topological properties.
 However, so far not even the simplest magnetic phase transition from AFM to FM order has been addressed in the context of accompanying changes of the non-trivial bulk topology~\cite{Cherifi}. Moreover, the realization of diverse magnetic topological phases on a square lattice is very challenging, which obstructs the observation of topological phase transitions driven by the changes in the respective magnetic structure of topologically non-trivial AFM systems considered so far~\cite{Mong,CFang,PZhou,JZhou,Tang,ZWang}. In sharp contrast to the case of the square lattice, the two-dimensional honeycomb structure as realized in graphene or silicene is known to naturally host a rich spectrum of quantum Hall effects~\cite{Kane,Haldane,chang,DXiao,Mak}, constituting thus the most favorable platform to uncover the complex interplay between global band topology and magnetic order. Furthermore, the strong geometric frustration of the bipartite honeycomb lattice makes the system prone to magnetic phase transitions as a function of different parameters, and can ultimately lead to the formation of various complex magnetic states.

In this work, based on tight-binding arguments and density functional theory calculations, we predict the emergence of the QSH effect in two-dimensional collinear antiferromagnets with honeycomb lattice structure. Using functionalized tin films $X$-Sn (where $X$ = H, F, Cl, Br, and I) as prototypical examples, we demonstrate that strain-induced magnetic phase transitions in these systems are accompanied by topological phase transitions from the QSH to the QAH insulator, as evidenced by the calculations of the spin and anomalous Hall conductivities, as well as by the analysis of the corresponding gapless edge states. Our findings are crucial for the realization of advanced concepts in topological antiferromagnetic spintronics, and their implementation within ``conventional" topological schemes associated up to date exclusively with ferromagnets and non-magnetic materials.

\begin{figure}
\centering
\includegraphics{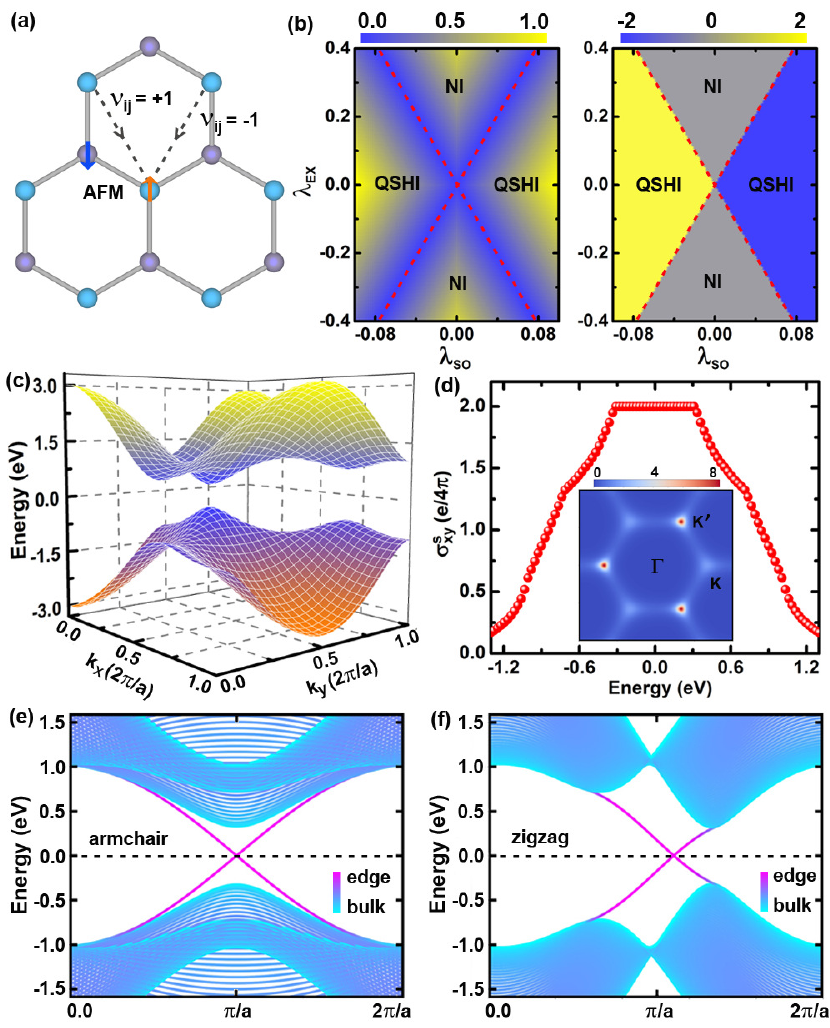}
\caption{ (a)~Sketch of the tight-binding model of a honeycomb antiferromagnet. (b)~Phase diagrams with respect to spin-orbit coupling $\lambda_{SO}$ and antiferromagnetic exchange $\lambda_{EX}$, where colors represent the bulk band gap (left) and the spin Hall conductivity $\sigma_{xy}^{S}$ quantized in units of $e/(4\pi)$ (right). (c)~Band structure of the antiferromagnetic quantum spin Hall insulator with $\lambda_{SO} = -0.1$ and $\lambda_{EX} = 0.2$. (d)~Corresponding energy dependence of $\sigma_{xy}^{S}$, with $\bf k$-space spin Berry curvature shown in the inset. (e,f)~Dispersion of the edge states, which reveal the topologically non-trivial nature of the model.}
\label{model}
\end{figure}

In two dimensions, as sketched in Fig.~\ref{model}(a), the intrinsic valley degree of freedom of the honeycomb lattice can be coupled to AFM order~\cite{XLi}, which breaks time-reversal symmetry protecting usually the QSH state~\cite{Hasan,Qi1}. Although the total magnetization vanishes in such a case, the electronic states can be spin-polarized leading to the realization of the QAH effect in antiferromagnets as predicted recently~\cite{PZhou}. However, if the combined symmetry $\mathcal P \mathcal T$ of time reversal $\mathcal T$ and spatial inversion $\mathcal P$ is preserved, each energy band is guaranteed to be doubly degenerate and not spin-polarized. As we show in the following, the $\mathcal P \mathcal T$ symmetry satisfied by collinear antiferromagnets with honeycomb structure (see Fig.~\ref{model}(a)) can manifest in the QSH effect. We begin by considering the  four-band AFM tight-binding model
\begin{equation}
\begin{split}
H =& -t\sum_{\langle ij \rangle\alpha} c^{\dagger}_{i\alpha}c^{\phantom{\dagger}}_{j\alpha}+i\lambda_{SO}\sum_{\langle\langle ij \rangle\rangle\alpha\beta}\nu^{\phantom{\dagger}}_{ij}\sigma^{z}_{\alpha\beta}c^{\dagger}_{i\alpha}c^{\phantom{\dagger}}_{j\beta}\\
 &+\lambda_{EX}\sum_{i\alpha}\mu^{\phantom{\dagger}}_{i}\sigma^{z{\phantom{\dagger}}}_{\alpha\alpha}c^{\dagger}_{i\alpha}c^{\phantom{\dagger}}_{i\alpha} \, ,
\label{eq:tbmodel}
\end{split}
\end{equation}
where $c^{\dagger}_{i\alpha}$($c^{\phantom{\dagger}}_{i\alpha}$) creates (annihilates) an electron with spin $\alpha$ on site $i$, $\langle ... \rangle$ and $\langle\langle ... \rangle\rangle$ restrict the sums to nearest and next-nearest neighbors, respectively, and $\sigma^z$ is a Pauli matrix  corresponding to the out-of-plane direction $z$. Apart from the usual hopping with amplitude $t$, Eq.~\eqref{eq:tbmodel} accounts for an intrinsic spin-orbit coupling ($\lambda_{SO}$) with $\nu_{ij}=\pm 1$ chosen as in Fig.~\ref{model}(a), and an AFM exchange ($\lambda_{EX}$), where $\mu_i=\pm 1$ has opposite values on the two sublattices. In the following, we quantify both $\lambda_{SO}$ and $\lambda_{EX}$ in units of $t$.
After diagonalizing the above bulk Hamiltonian in reciprocal space, we present in Fig.~\ref{model}(b) the resulting evolution of the tunable global band gap throughout the phase space spanned by $\lambda_{SO}$ and $\lambda_{EX}$. Apparently, the electronic structure of the model is always insulating at half-filling (see also Fig.~\ref{model}(c)) except along the lines in the phase space where $\lambda_{EX}=\pm 3\sqrt{3} \lambda_{SO}$, at which the band gap closes at $K$ and $K^\prime$, respectively. 

We prove now by explicit calculation of the corresponding topological invariant that the zero-gap lines in the phase diagram of Fig.~\ref{model}(b) separate the trivial insulating from the topologically distinct QSH phase of the antiferromagnet. In order to identify uniquely different topological phases of the $\mathcal P\mathcal T$-symmetric model, we analyze the spin Hall conductivity $\sigma_{xy}^{S}$ that can be obtained from the Kubo formula~\cite{Sinova}
\begin{equation}
\sigma_{xy}^{S}=e\hbar\int\frac{d^2 k}{(2\pi)^2}\, \Omega^{S}({\bf k}) \, ,
\end{equation}
with the spin Berry curvature of all occupied states
\begin{equation}
 \Omega^{S}({\bf k})=-2{\rm Im} \sum_{m\ne n}\frac{\bra{\psi_{m{\bf k}}}J_{x}^{s}\ket{\psi_{n {\bf k}}}\bra{\psi_{n{\bf k}}}\upsilon_{y}\ket{\psi_{m {\bf k}}}}{(E_{n\bf k}-E_{m\bf k})^2} \, .
\end{equation}
Here, $\upsilon_{i}$ is the $i$th Cartesian component of the velocity operator, $|\psi_{n\bf k}\rangle$ is an eigenstate of the lattice-periodic Hamiltonian with energy $E_{n\bf k}$, and $J_{x}^{s}=(\hbar/2)\{\sigma^z, \upsilon_x\}$ describes a spin current flowing into $x$ direction with spin polarization perpendicular to the plane. Figure~\ref{model}(b) displays the phase-space map of the quantized spin Hall conductivity, revealing the complex interplay of spin-orbit coupling and AFM exchange interaction in establishing different insulating phases of the model. By engineering the electronic structure of the $\mathcal P\mathcal T$-symmetric antiferromagnet, we can realize either the trivial insulator ($\sigma_{xy}^S=0$) or the topologically non-trivial QSH insulator distinguished by a finite spin Hall conductivity $\sigma_{xy}^S=\pm 2 e/(4\pi)$. We confirm these findings by computing the spin Chern number $\mathcal C_S$ in the respective insulating regions, allowing for the alternative representation $\sigma_{xy}^S=\mathcal C_S e/(2\pi)$~\cite{Yang,Prodan}.

Apart from the quantization of spin Hall conductivity and spin Chern number, an exceptional feature of QSH insulators is the emergence of metallic edge states connecting bulk valence and conduction bands. To observe these peculiar states originating from the global topology, we calculate the dispersion of one-dimensional ribbons with armchair or zigzag edges as shown in Figs.~\ref{model}(e) and~\ref{model}(f). Using as an example the parameters $\lambda_{SO} = -0.1$ and $\lambda_{EX} = 0.2$ to reside in the QSH region (cf. Fig.~\ref{model}(b)), we clearly perceive the appearance of gapless helical states located at either of the sample edges. This constitutes the final proof of the QSH state of the considered collinear antiferromagnetic model on a honeycomb lattice preserving $\mathcal P \mathcal T$ symmetry.

\begin{figure*}
\centering
\includegraphics{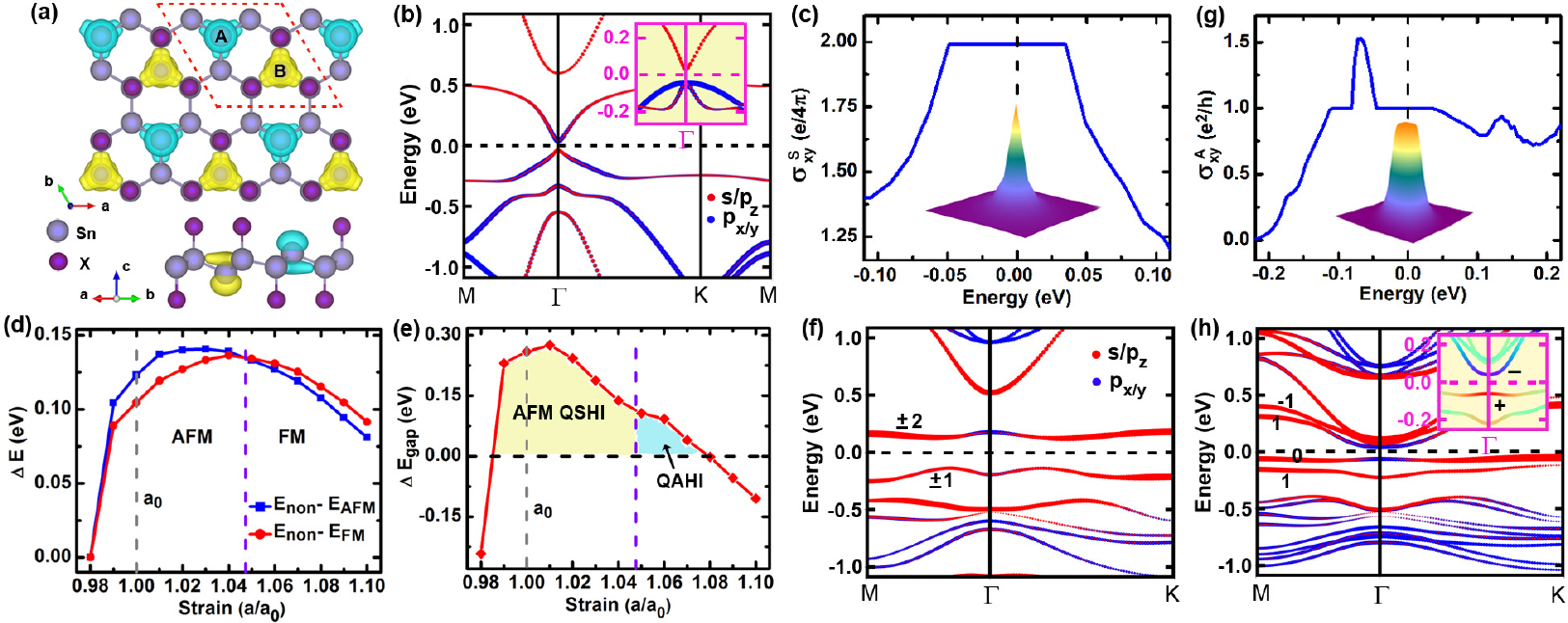}
\caption{ (a)~Crystal structure of functionalized tin films $X$-Sn ($X$ = H, F, Cl, Br, and I), with the antiferromagnetic spin-density distribution in the case of H-Sn on top. Red dashed lines denote the unit cell, and defect sites without $X$ are labeled as A and B. (b) Orbitally resolved band structures of H-Sn weighted by the Sn-$s,p_z$ and Sn-$p_x,p_y$ orbital characters, either without (inset) or with spin-orbit coupling. The energy dependence of the spin Hall conductivity $\sigma_{xy}^{S}$ is shown in (c) with a quantized value, which arises mainly from the spin Berry curvature near $\Gamma$ (inset), in the spin-orbit coupling gap. (d) Energy differences of ferromagnetic (FM) and antiferromagnetic (AFM) configuration in I-Sn, and (e) corresponding ground-state band gap as a function of the applied strain $a/a_0$ driving both magnetic and topological phase transitions. Orbitally resolved band structures in (f) quantum spin Hall phase and (h) quantum anomalous Hall phase under 5\% tensile strain with spin-orbit coupling. (g) Energy dependence of the anomalous Hall conductivity $\sigma_{xy}^{A}$ in the strained case, exhibiting a quantized value throughout the bulk band gap originating from a prominent Berry curvature near $\Gamma$ (inset). The numbers in (f) and (h) label Chern numbers for the bands around the Fermi level, and the inset in (h) presents the expectation value of angular momentum $L_z$ as colors ranging from dark red (positive, ``$+$") to dark blue (negative, ``$-$").}
\label{snh}
\end{figure*}

Having established the general concept of a topologically non-trivial honeycomb AFM insulator from a model, we aim now at the prediction of realistic antiferromagnets that exhibit the anticipated effect. In particular, we focus on honeycomb systems of functionalized tin films $X$-Sn (where $X$= H, F, Cl, Br, and I), Fig.~\ref{snh}(a), motivated by the fact that Sn with graphene-like structure has been successfully fabricated~\cite{Zhu} and was predicted to transform into a non-magnetic QSH insulator~\cite{Xu} or a ferromagnetic QAH insulator~\cite{wu} depending on the functionalization. We perform first-principles calculations of the electronic structure of $X$-Sn using the generalized gradient approximation to density functional theory as implemented in the Vienna ab initio simulation package VASP~\cite{Kresse,Kresse1} and the \texttt{FLEUR} code~\cite{fleur}. Subsequently, we construct maximally-localized Wannier functions employing \textsc{wannier}{\footnotesize{90}} in conjunction with the \texttt{FLEUR} package to compute the Berry curvatures and edge states~\cite{Mostofi,Freimuth,supp}.

At first, we address the question whether functionalized Sn films can develop ground-state AFM order under proper electronic-structure engineering. Indeed, starting from the fully decorated case, we find that antiferromagnetism is induced when $X$ vacancies in the $2$$\times$$2$ corrugated honeycomb geometry are considered, Fig.~\ref{snh}(a), since the total energy in this case is lowered by several meV as compared to the FM configuration (see $E_m$ in Table~\ref{default}). Considering as an example H-Sn, we further reveal in Fig.~\ref{snh}(a) the distribution of the ground-state spin density in this system, which is centered but opposite in spin character around the two Sn sites with H-vacancy. While the total density of states of H-Sn is symmetric with respect to spin direction, the partial density of states projected onto Sn$_A$ and Sn$_B$ is spin-polarized~\cite{supp}. The latter projections are identical except for a global sign, leading to total spin moments of $\pm 1\,\mu_B$ on the vacancy sites. Analogously, all other systems $X$-Sn considered here develop an AFM order that can couple to the valley degree of freedom of the honeycomb structure.

To uncover first hallmarks of the topologically distinct QSH phase in the $\mathcal P\mathcal T$-symmetric antiferromagnet H-Sn, we present in Fig.~\ref{snh}(b) orbitally-resolved band structures highlighting an $s$-$p$ band inversion at the $\Gamma$-point upon including the effect of spin-orbit interaction. In this case, the respective global band gap amounts to $\Delta E_{gap}=78\,$meV. Following the dependence of the spin Hall conductivity on the band filling, Fig.~\ref{snh}(c), we clearly demonstrate the non-trivial band topology in H-Sn as indicated by the quantization of $\sigma_{xy}^S$ within the insulating region. As summarized in Table~\ref{default}, the QSH effect emerges in all of the studied collinear antiferromagnets with $X=$ H, F, Cl, Br, and I, with the record band gap of $271\,$meV realized in Br-Sn. This suggests the high stability of this topological AFM phase and the possibility of its experimental observation in this family of compounds.

The manifestation of the QSH effect in the considered antiferromagnets offers an intriguing platform to scrutinize the complex interplay of topological phases with magnetic order as well as their external control, which is appealing from the viewpoint of both fundamental research and practical applications. In the following, we focus on the application of strain that proved to be highly efficient in tailoring magnetic and topological properties~\cite{Cherifi,niutlse}. We quantify the magnitude of strain by the ratio $a/a_0$, where $a_0$ and $a$ denote the in-plane lattice constants of the unstrained and strained crystals, respectively. Using I-Sn as an example, we show in Fig.~\ref{snh}(d) the energy difference of FM and AFM configuration revealing a qualitatively distinct behavior of the magnetic order with respect to either compressive or tensile strain. While the unstrained system is an insulating antiferromagnet as discussed before, under compressive strain I-Sn turns into a non-magnetic metal as apparent from the evolution of the bulk band gap in Fig.~\ref{snh}(e). However, if we apply critical tensile strain of about $5\%$, the material experiences a phase transition from AFM order to a ferromagnetic state with non-zero energy gap.

 \begin{table}
\caption{Relaxed lattice constant $a_{0}$ and energy difference $E_{m}=E_{FM}-E_{AFM}$ at this lattice constant. All unstrained compounds $X$-Sn are antiferromagnetic quantum spin Hall insulators with the band gap $\Delta E_{gap}$. Under the critical strain $a_{c}/a_0$, the system undergoes a phase transition to a ferromagnetic quantum anomalous Hall insulator with the band gap $\Delta E_{c}$. The values of the quantized spin Hall and anomalous Hall conductivities in units of $e/(4\pi)$ and $e^2/h$, respectively, are shown. }
\begin{center}
\begin{tabular}{ c c c c c c}
\hline
\hline
 compound   & ~H-Sn~&~F-Sn~ & ~Cl-Sn~ & ~Br-Sn~ & ~I-Sn~   \\
\hline
$a_{0}$ (\AA)  & 9.46& 9.68 & 9.65 & 9.63 & 9.61  \\
$E_{m}$ (meV)  & 10.9 &5.9 & 9.0 & 10.2 & 18.6  \\
$\Delta E_{gap}$ (meV)   & 78 & 258 & 254 & 271 &  259  \\
$a_{c}/a_0$ & 9\% & 3\%  & 3\% & 4\% & 5\% \\ 
$\Delta E_{c}$ (meV)   & -- & 140 & 134 & 126 &  93  \\
$\sigma_{xy}^{S}\,|\,\sigma_{xy}^{A}$       & 2 $|$ -- & 2 $|$ 1 & 2 $|$ 1 & 2 $|$ 1 &  2 $|$ 1   \\
\hline
\hline
\end{tabular}
\end{center}
\label{default}
\end{table}

In order to explore the possibility of a topological phase transition accompanying this strain-driven drastic change of the magnetic properties in I-Sn, we present in Figs.~\ref{snh}(f) and~\ref{snh}(h) the electronic band structures in the antiferromagnetically and ferromagnetically ordered ground state upon taking into account the effect of spin-orbit interaction. Each of the energy bands in the AFM ground state is doubly degenerate with opposite Chern numbers, leading ultimately to the non-trivial QSH phase in the honeycomb antiferromagnets. Under a critical tensile strain of $5\%$, the ground state becomes ferromagnetic with a large global energy gap of $\Delta E_{c}=93\,$meV induced by spin-orbit coupling, which renders an experimental observation generally feasible. While the antiferromagnet hosts predominantly $p_z$ states around the gap, in the strained ferromagnet orbitally-purified bands of $p_x \pm i p_y$ character (cf. Fig.~\ref{snh}(h)) are brought into the close vicinity of the Fermi level as a result of FM ordering. This orbital purification is expected to naturally give rise to the non-zero Chern number~\cite{Zhang}. Indeed, following in Fig.~\ref{snh}(g) the energy dependence of the anomalous Hall conductivity $\sigma_{xy}^{A} = e^2/(2\pi h) \int\Omega({\bf k})d^2k$, where $\Omega(\bf k)$ is the momentum Berry curvature of all occupied states, we note the quantization of $\sigma_{xy}^A$ as unique fingerprint of the QAH phase emerging in the strained insulating ferromagnet. Surprisingly, also the spin Hall conductivity acquires a finite quantized value of $e/(4\pi)$, indicating the coexistence of both QSH and QAH phases in I-Sn, to which we may thus refer as quantum spin Chern insulator~\cite{Zhang}. As outlined in Table~\ref{default} and Ref.~\cite{supp}, the topological phase transition from the QSH to the QAH insulator coming with the aforementioned magnetic transition occurs in all considered compounds, even with the tensile strain as small as 3\% and the band gap $\Delta E_c$ as large as 140 meV, except for H-Sn, which becomes metallic under tensile strain.

\begin{figure}
\centering
\includegraphics{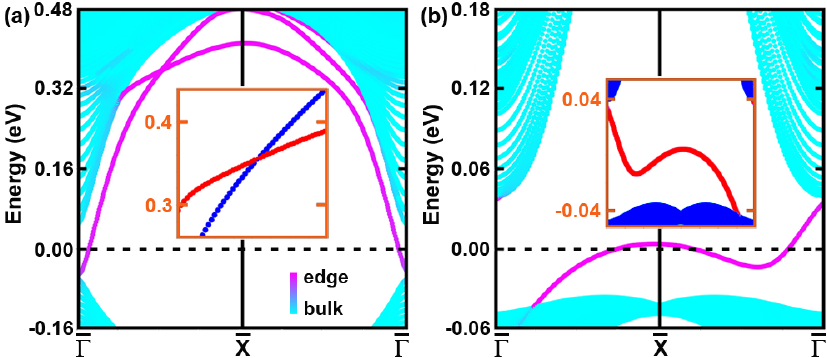}
\caption{Band structures of one-dimensional zigzag ribbons for (a) antiferromagnetic H-Sn in the quantum spin Hall phase, and (b) ferromagnetic I-Sn under 5\% tensile strain exhibiting the quantum anomalous Hall effect. Helical and, respectively, chiral edge states are visible in the bulk band gap. Insets show the spin polarization of the edge states (a), or the gapless states at the opposite edge (b).}
\label{edge}
\end{figure}

Eventually, we complement our topological characterization of the magnetic QSH and QAH insulators by addressing explicitly their metallic edge states that are of helical and chiral nature, respectively. Figure~\ref{edge}(a) reveals the appearance of pairs of spin-polarized gapless edge states at either of the two boundaries of a one-dimensional zigzag ribbon constructed from the unstrained QSH antiferromagnet H-Sn. In contrast, the exotic edge states of the strained QAH insulator I-Sn, as shown in Fig.~\ref{edge}(b), are chiral in the sense of opposite group velocity at opposite edges.

In summary, we demonstrated the emergence of the quantum spin Hall effect in collinear honeycomb antiferromagnets, which obey the combined $\mathcal P \mathcal T$ symmetry, by explicit calculation of the respective topological invariant and a detailed analysis of the edge states. We further identified functionalized tin films as promising material candidates with complex topology that are in principle experimentally accessible. Remarkably, we discovered that these systems undergo a topological phase transition to the quantum anomalous Hall regime accompanying the strain-induced magnetic transition. The presented results not only advance our general understanding of the quantum Hall effects and phase transitions but also put forward potential applications of topological matter in antiferromagnetic spintronics. In particular, our discovery of possible material systems which reside in the vicinity of magnetic and topological transition points opens an intriguing possibility of harvesting the benefits of ferromagnetic and antiferromagnetic topological phases within the same device.   

This work was supported by the Priority Program 1666 of the German Research Foundation (DFG) and the Virtual Institute for Topological Insulators (VITI). We acknowledge computing time on the supercomputers JUQUEEN and JURECA at J\"{u}lich Supercomputing Centre and JARA-HPC of RWTH Aachen University.

\end{document}